\documentstyle[stwol,PRL,page1,epsf]{article}

\def \egev    {GeV}   
\def \mgev    {GeV/$c^{2}$}   
\def \mtev    {TeV/$c^{2}$}   
\def \pgev    {GeV/$c$}

\def \et      {${\rm E_{T}}$ }
\def \pt      {${\rm p_{T}}$ }
\def \miset   {$\,/\!\!\!\!E_{T}$ }

\def \z0      {$Z^{0}$}
\def \ppbar   {$p\bar{p}$ }
\def \etal    {{\it et.al.}}

\begin{document}

\title{
\vspace*{1.8cm}
       SEARCH FOR LEPTOQUARKS AT CDF \\
      {\small (To appear in the Proceedings of the XII
     Hadron Collider Physics Symposium, June5-11 1997, Stony brook)}
}       

\author{ Hisanori Sam Kambara \\
        (The CDF Collaboration) }

\address{
Department de Physique Nucl\'{e}aire et Corpusculaire \\
Universit\'{e} de Gen\`{e}ve, CH-1211 Gen\`{e}ve 23, Switzerland}

\twocolumn[\maketitle\abstracts{ We present the result of
  direct leptoquark searches based on 110 $pb^{-1}$ of
  integrated luminosity collected by the Collider Detector
  at Fermilab during the 1992-93 and 1994-95 Tevatron runs
  at $\sqrt{s}=1.8$ TeV.  We present upper limits on the
  production cross sections as a function of the leptoquark
  mass.  Using the NLO calculation of the leptoquark-pair
  production cross sections we extract lower mass limits for
  1$^{st}$, 2$^{nd}$, and 3$^{rd}$ generation leptoquarks.
  We also present the result of an indirect search for
  Pati-Salam leptoquarks via exclusive $e\mu$ decay modes of
  $B^{0}_{s}$ and $B^{0}_{d}$.}]

\section{Introduction}

The Standard Model which is based on the strong and
electroweak interactions with the $SU(3) \otimes SU(2) \otimes
U(1)$ gauge group has been successful in describing the
phenomenology of high energy particle physics. Features of
the Standard Model such as the mass spectrum of the three
fermion generations or the quark-lepton symmetry are not
yet understood.
                                                  
Leptoquarks appear in several extensions to the Standard
Model.  They are color-triplet bosons which mediate
interactions between quarks and leptons.  Leptoquarks with a
mass accessible through direct production at the current
accelerators are usually assumed to couple to quarks and
leptons of the same generation~\cite{buckmuller}, in order
to avoid large flavour-changing neutral current processes.
One therefore speaks of leptoquarks of first, second, or
third generation, which we generically denote by $\Phi_i$,
$i=1,2,3$. Quantum numbers such as the charge $Q$ and weak
isospin are model dependent.

For very heavy leptoquarks, well above the TeV scale, FCNC
constraints can allow couplings to quarks and leptons in
different generations. This is the case, for example, of the
so-called Pati-Salam leptoquarks~\cite{patisalam}.  They
appear as gauge vector bosons in a grand-unified extension
of the Standard Model based on an enlarged color group
$SU(4)_{c}$, which contains the lepton number as fourth
color.  A color-triplet set of $SU(4)$ gauge bosons acquires
a mass when the $SU(4)$ group is broken, at a large mass
scale, to $SU(3)\times U(1)$.  The gauge nature of these
leptoquarks dictates that they should couple to all
generations, thereby inducing, among other processes, decays
such as $B^0_s\rightarrow e\mu$ and $B^0_d\rightarrow e
\mu$~\cite{scott}.  Setting limits on the branching ratio of
these otherwise forbidden processes can probe masses in the
multi-TeV range.

We present in this work the preliminary results of
leptoquark searches performed by CDF using the full 110
pb$^{-1}$ of integrated luminosity from the Run IA+B data
samples collected at $\sqrt{s} =1.8$ TeV during the
1992-1995 Tevatron run.  The direct searches are discussed
in section 2, and the indirect search using
$B^0_s\rightarrow e\mu$ and $B^0_d\rightarrow e \mu$ is
presented in section 3.

\section{Direct Search for Pair Produced Leptoquarks}

Leptoquarks can be produced in pairs in \ppbar collisions
via strong interactions, through gluon-gluon fusion and
$q\bar q$ annihilation~\cite{hewett}.  The contribution to
the production rate from the direct $\Phi \bar q l$ coupling
is suppressed relative to the dominant QCD mechanisms.  The
cross section can therefore be calculated independently of
the value of the leptoquark coupling $\lambda$, and is
currently evaluated up to next-to-leading order (NLO)
accuracy~\cite{lqxsectheo}.
                           
In this study we concentrate on leptoquarks which can decay
to a quark and a charged lepton, with a non-zero branching
ratio $\beta$.  The search is therefore based on events with
two charged leptons plus $\ge 2$ jets. The jets are defined
by the standard cone algorithm using the cone of 0.7 on
$\eta-\phi$ space, and, unless otherwise stated, are
required to be within $|\eta| < 2.4 $.
                                  
\subsection{$3^{rd}$ generation search}
CDF has searched for third generation leptoquarks decaying
into two $\tau$'s and two jets ($\Phi_3\bar{\Phi}_{3}
\rightarrow \tau^{+} \tau^{-} j j$).  The results of this
search have been published in ref.~\cite{lq3cdf}, to which
we refer the reader for full details.  We present here a
brief summary of the analysis.  We require one $\tau$ to
decay leptonically, the other hadronically.  In the first
case we consider $e$ or $\mu$ decays, with the following
selection criteria: \pt$(e,\mu) > 20$ \pgev, the \miset
should point within 50$^{0}$ of the lepton direction and the
leptons should be isolated.  For the hadronic $\tau$ decay
we require 1 or 3 charged tracks within a 10$^{0}$ cone
about the jet axis and no other tracks above 1 \pgev \,
between the 10$^{0}$ and 30$^{0}$ cones.  Having selected
$\tau\tau$ events we then require two additional jets with
\et $ > 20 $ \egev.  The jet algorithm with a 0.4 cone in
$\phi-\eta$ space is used, and b-tagging is not required.

\begin{figure}[tb]
\center
  \vspace*{-2.5cm}
  \hspace*{-2.0cm}
  \epsfxsize=11cm
  \epsfysize=13cm
  \epsffile{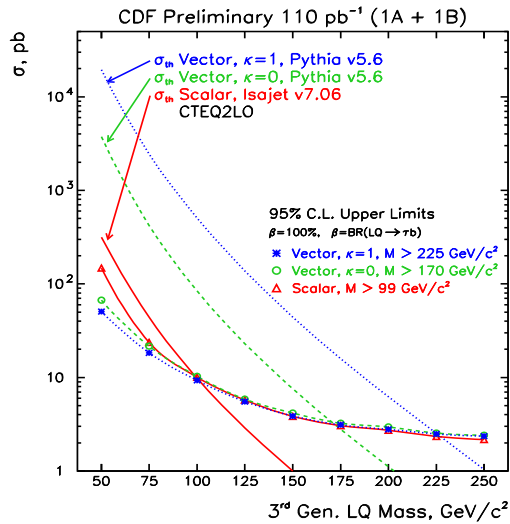}
  \vspace{-3.5cm}
  \caption{\small 
    95 \% C.L. CDF cross section limit for $\Phi_3$.}
  \label{fig_lq3}                              
\end{figure}
The final candidate events are selected by rejecting the
events with the $\tau\tau$ invariant mass consistent with
the $Z^{0}$ mass.  Events where the primary lepton and the
leading tau-jet track have an invariant mass in the range 70
to 110 \mgev are therefore removed.  After the event
selection, we observe one event as a $\Phi_3$ candidate,
with an expected background of $2.4^{+1.2}_{-0.6}$, coming
mainly from $(Z^{0} \rightarrow \tau^{+}\tau^{-}) +
jets$.  Accounting for the selection efficiency, we can then
exclude $ M_{\Phi_3} < 99 $ \mgev at 95\% C.L. For vector
leptoquarks with anomalous chromomagnetic moments
parameterised by $\kappa$~\cite{Bluemlein96}, assuming
$\beta$ = 1, our new limits exclude $M_{\Phi_3} < 170$
\mgev\, and $M_{\Phi_3} < 225 $ \mgev\, for $\kappa=0$ and
$\kappa=1$ respectively.  The results are summarised in
Figure~\ref{fig_lq3}.

\subsection{$2^{nd}$ generation search}
The search for $\Phi_2$ production looks for events where
the $\Phi_2$ pair decays into dimuon + dijets ($\Phi_2\Phi_2
\rightarrow \mu^{+} \mu^{-} j j$).  A previous CDF
study~\cite{lq2cdf} excluded $M_{\Phi_2} < 131(96) $ \mgev\,
for $\beta = 1.0(0.5)$ using 19 pb$^{-1}$ CDF data from Run
IA.  Results have also been published by
D$\emptyset$~\cite{lq2d0}.  Here we present the new CDF
limit using 110 pb$^{-1}$ of data.
                                                              
\begin{figure}[tb]
\center
  \hspace*{-0.5cm}
  \epsfysize=8.5cm
  \epsffile{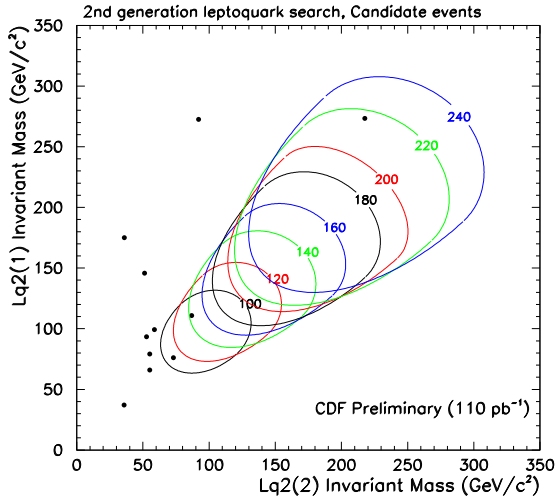}
  \vspace{-1.0cm}
  \caption{\small $M_{\mu j}$ distribution for events 
           before the mass balancing cut.} 
 \label{fig_lq2_dist}
\end{figure}
We require events with two muons, satisfying \pt $ > 30$ \pgev
$(\mu_{1})$ and $ > 20 $ \pgev $(\mu_{2})$.  Muon quality
cuts, such as isolation, a fiducial requirement and vertex
matching, are also applied.  Then we ask for $\geq$ 2 jets
with \et$^{(1)} > 30 $ \egev\, and \et$^{(2)} > 15 $ \egev.
The $Z^{0}$ and other resonances are removed by rejecting
events with a dimuon invariant mass in the regions
$M_{\mu\mu} < 10 $ \mgev and $76 < M_{\mu\mu} < 106 $
\mgev).  A total of 11 events pass the above selection cuts
and are shown in Figure~\ref{fig_lq2_dist}, plotted in the
muon-jet invariant-mass plane ($M_{\mu j}^1 $ v.s.  $M_{\mu
  j}^2$).  Of the two possible muon-jet pairings we choose
the one with the smallest invariant mass difference.

In the case of leptoquark-pair decays the two muon-jet
systems have approximately the same mass, within the mass
resolution $\sigma$.  We therefore search for leptoquark
candidates by selecting events in a $3\sigma$ region of the
$M_{\mu j}^1$ vs. $M_{\mu j}^2$ plane around any given mass,
as shown in Fig.~\ref{fig_lq2_dist}.  This requirement
reduces the background substantially, since in the
background events, the reconstructed muon-jet invariant
masses are not correlated.  Possible background sources are
mainly from Drell-Yan and heavy flavour production and
decay. The total signal detection efficiency for the signal
depends on the $\Phi_2$ mass, and it is calculated to be
15\% at $M_{\Phi_2}= 200$ \mgev.  The major source of
systematic uncertainty on the efficiency comes from the
effects of gluon radiation.  We compute the experimental
cross section limit with a 20 \% systematic uncertainty.

\begin{figure}[tb]
\center
  \hspace*{-0.5cm} 
  \epsfysize=8.5cm
  \epsffile{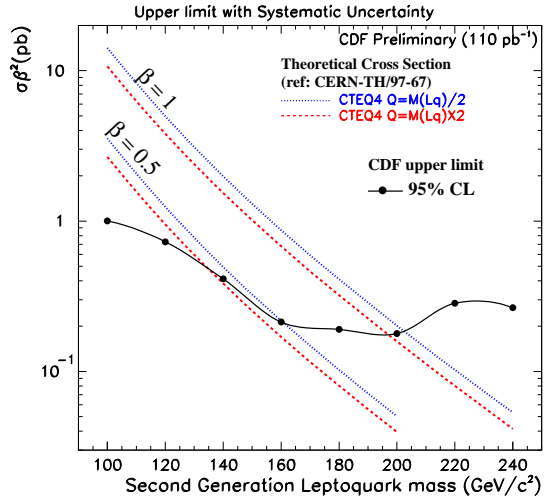} 
  \vspace{-1.0cm}
  \caption{\small 95\% C.L. CDF cross section limit for $\Phi_2$.}
  \label{fig_lq2_limit}                                         
\end{figure}
Figure~\ref{fig_lq2_limit} shows the CDF 95\% C.L.
cross-section limit on the $\sigma(p\bar p \rightarrow
\Phi_2\bar{\Phi}_2) \beta^{2}$.  Comparing to the NLO
cross-section calculation~\cite{lqxsectheo} a limit of
$M_{\Phi_2} > 195$ \mgev\, ($\beta = 1.0$) is derived.
                                           
\subsection{$1^{st}$ generation search}
The previous search for $\Phi_1$ at CDF set a mass limit of
$113(80)$ \mgev\, for $\beta=1.0(0.5)$, using 4.05 pb$^{-1}$
CDF data~\cite{lq1cdf}.  Interest in extending this search
to the mass region around 200~\mgev/, is stimulated by the
recent results from the HERA experiments, reporting an
excess of high-$Q^2$ deep inelastic scattering
events~\cite{lq1hera}.
                 
\begin{figure}[tb]
\center
  \hspace*{-0.5cm} 
  \epsfysize=8.5cm
  \epsffile{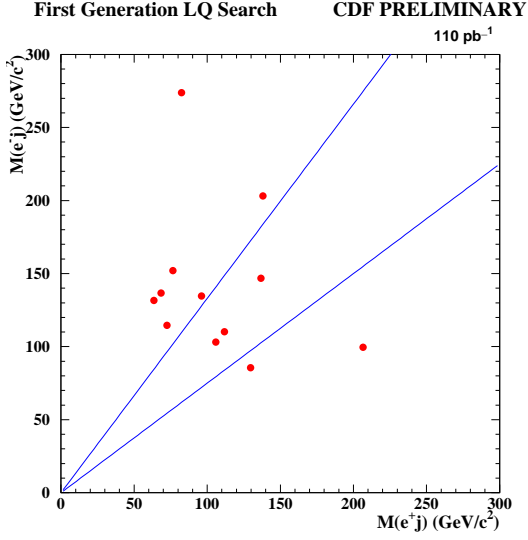} 
  \vspace{-1.0cm} 
  \caption{\small $M_{ej}$ distribution of the events before
                  leptoquark mass constraint.}
  \label{fig_lq1_dist}
\end{figure}
The event selection for $\Phi_1$ is similar to the $\Phi_2$
selection, requiring high-\et dielectrons in an event with
at least two jets.  We first select two electrons by
requiring \et $ > 25 $ \egev.  We then ask for two or more
jets in the whole rapidity range with \et$(j_{1}) > 30$
\egev\, and \et$(j_{2}) > 15 $ \egev.  A $Z^{0}$ veto is
applied to the dielectron mass (76 \mgev $ < M_{ee}$ 106
\mgev).  An additional cut is applied by requiring minimum
values for the transverse-energy sum of the dielectron and
dijet systems: $E_{T}(e_{1}) + E_{T}(e_{2}) > 70 $ \egev\,
and $E_{T}(j_{1}) + E_{T}(j_{2}) > 70 $ \egev.  This cut
($\Sigma E_{T}$ cut) is efficient in removing major
backgrounds such as Drell-Yan.  The events passing this cut
are displayed on the $M_{ej}^1$ vs.  $M_{ej}^2$ plane in
Figure~\ref{fig_lq1_dist}.  The $e$-jet pairings are chosen
in the same way as in the $\Phi_2$ search.

\begin{figure}[tb]
\center
  \hspace*{-0.5cm} 
  \epsfysize=8.5cm
  \epsffile{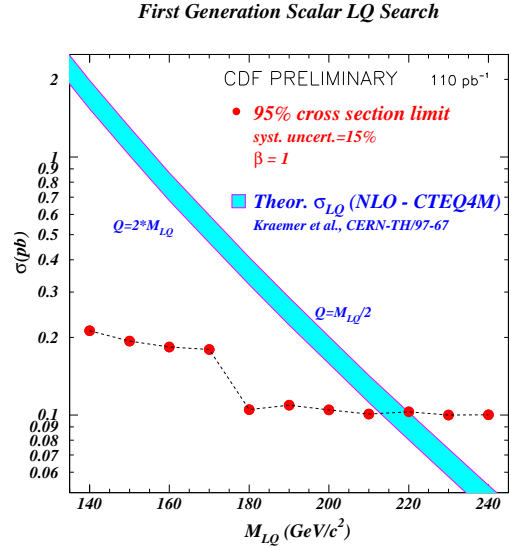} 
  \vspace{-1.0cm} 
  \caption{\small
       95 \% C.L. CDF cross section limit for $\Phi_1$.}
  \label{fig_lq1_limit}                                
\end{figure}
We select the $\Phi_1$ candidates by requiring the two
$M_{ej}$'s in the event to be within $<0.2\times M_{\Phi_1}$.
The final $\Phi_1$ candidates for a given mass $M$ are
selected by choosing events with the mean $M_{ej}$ of the
pair to be within $3\sigma$ of $M$.

The signal acceptance is evaluated using the PYTHIA Monte
Carlo.  It varies between 21\% ($M_{\Phi_1}=$140 \mgev) and
28\% ($M_{\Phi_1}=$240 \mgev).  A 15\% systematic
uncertainty is used to compute the 95\% C.L. CDF $\Phi_1$
cross section limit, which is shown in
Figure~\ref{fig_lq1_limit}, compared with the theoretical
calculations~\cite{lqxsectheo}.  From this, we derive a
limit of $M_{\Phi_1} > 210 $ \mgev\, for $\beta=1.0$.
                             
\section{Indirect Search for Leptoquarks 
(Search for the decays $B^0_s\rightarrow e \mu$ and 
$B^0_d\rightarrow e \mu$)} 

CDF has searched for the decays $B^0_s\rightarrow e \mu$ and
$B^0_d\rightarrow e \mu$ using $\approx$ 85 $pb^{-1}$ of Run
IB data. We select events with an oppositely charged
$e\mu$-pair, the electron with $E_T >$ 5.0 GeV and the muon
with $P_T>$2.5 Gev/c. In addition we require the proper
decay length $c\tau$ of the $e\mu$ system to be larger than
$200 \mu m$ and that the reconstructed momentum vector of
the $e\mu$ pair point back to the primary vertex.  We find
no $B^0_d$ candidates in a mass window of $5.174 - 5.384$
\mgev ($\pm 3 \sigma$ of our mass resolution) and one
$B^0_s$ candidate in a mass window of $5.270 - 5.480$ \mgev.

\begin{figure}[tb]
\center
  \hspace*{-0.5cm}
  \epsfysize=8.5cm
  \epsffile{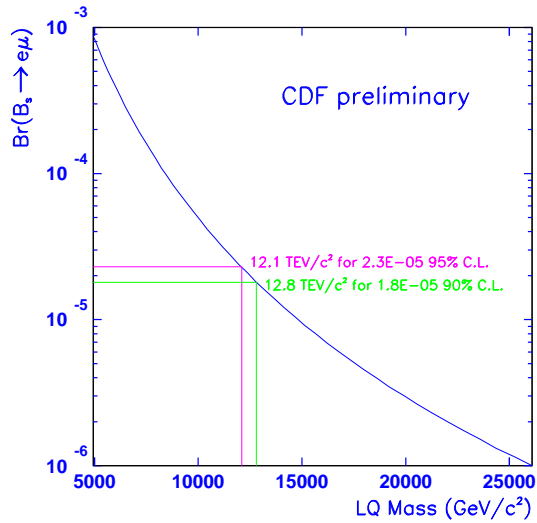}
  \vspace{-1.0cm}
  \caption{\small 
       Pati-Salam leptoquark mass limit.}
  \label{fig_b03}
\end{figure}
We extract 95\% CL limits at Br($B^0_s\rightarrow e \mu) <
2.3 \times 10^{-5}$ and Br($B^0_d\rightarrow e \mu) < 4.4
\times 10^{-6}$, including the systematic uncertainties.
From this we derive the following 95\% C.L. limits on the
mass of Pati-Salam leptoquarks as discussed in
ref.~\cite{scott}: $M>12.1$ TeV/$c^2$ for the $B_s$ (as
shown in Fig.~\ref{fig_b03}) and $M>18.3$ TeV/$c^2$ for the
$B_d$.  This last limit improves the CLEO bound of
16~TeV/$c^2$~\cite{cleo}.

\section{Summary}

Using the full Run IA+B CDF data at the Tevatron, we have
searched for direct production of leptoquark pairs in all
three generations. For pair-produced scalar leptoquarks, the
searches exclude $M_{\Phi_1} < 210 $ \mgev\, ($\beta=1$),
$M_{\Phi_2} < 195 $ \mgev\, for $\beta=1.0$, and $M_{\Phi_3}
< 99 $ \mgev\, for $\beta=1.0$.  For the $\Phi_3$ search, we
also set limits for vector leptoquarks, excluding
$M_{\Phi_3} < 225(170) $ \mgev\, for $\kappa=1(0)$.  CDF
also performed an indirect leptoquark search via $B^0_s
\rightarrow e\mu $ and $B^0_d \rightarrow e\mu $, setting
preliminary mass limits for Pati-Salam type leptoquarks at
12.1 \mtev\, from $B^0_s$ decays, and 18.3 \mtev\, from
$B^0_d$ decays.
                            
\vspace*{0.5cm}
{\bf Acknowledgements } \\

We thank the Fermilab staff and the technical staffs of the
participating institutions for their vital contributions.
This work was supported by the U.S. Department of Energy,
the National Science Foundation, the Istituto Nazionale di
Fisica Nucleare (Italy), the Ministry of Science, Culture
and Education of Japan, the Natural Sciences and Engineering
Research Council of Canada, the National Science Council of
the Republic of China, the A.~P.~Sloan Foundation and the
A.~von Humboldt-Stiftung.

\end{document}